\documentclass[conference]{IEEEtran}
\IEEEoverridecommandlockouts
\usepackage{cite}
\usepackage{amsmath,amssymb,amsfonts}
\usepackage{algorithmic}
\usepackage{graphicx}
\usepackage{textcomp}
\usepackage{xcolor}
\usepackage{multirow}
\usepackage{booktabs}
\usepackage{hyperref}
\usepackage{authblk}
\usepackage{subcaption}
\usepackage[font=small,labelfont=bf]{caption}
\hypersetup{
    colorlinks=true,
    linkcolor=blue,
    filecolor=magenta,      
    urlcolor=cyan,
    citecolor = magenta,
    pdfpagemode=FullScreen,
}

\makeatletter
\newcommand*\titleheader[1]{\gdef\@titleheader{#1}}
\AtBeginDocument{%
  \let\st@red@title\@title
  \def\@title{%
    \bgroup\normalfont\large\centering\@titleheader\par\egroup
    \vskip1em\st@red@title}
}
\makeatother

\def\BibTeX{{\rm B\kern-.05em{\sc i\kern-.025em b}\kern-.08em
    T\kern-.1667em\lower.7ex\hbox{E}\kern-.125emX}}

\title{
Circuit and System Technologies for Energy-Efficient Edge Robotics
}
\IEEEspecialpapernotice{(Invited Paper)}
\titleheader{2022 IEEE 27th Asia and South Pacific Design Automation Conference (ASP-DAC)}
\author[]{Zishen~Wan, Ashwin Sanjay Lele, Arijit~Raychowdhury\\School of Electrical and Computer Engineering, Georgia Institute of Technology, Atlanta, USA \\ zishenwan@gatech.edu, alele9@gatech.edu, arijit.raychowdhury@ece.gatech.edu }

\begin{document}


\maketitle

\begin{abstract}
As we march towards the age of ubiquitous intelligence, we note that AI and intelligence are progressively moving from the cloud to the edge. 
The success of Edge-AI is pivoted on innovative circuits and hardware that can enable inference and limited learning in resource-constrained edge autonomous systems.
This paper introduces a series of ultra-low-power accelerator and system designs on enabling the intelligence in edge robotic platforms, including reinforcement learning neuromorphic control, swarm intelligence, and simultaneous mapping and localization. We put an emphasis on the impact of the mixed-signal circuit, neuro-inspired computing system, benchmarking and software infrastructure, as well as algorithm-hardware co-design to realize the most energy-efficient Edge-AI ASICs for the next-generation intelligent and autonomous systems.

\end{abstract}

\section{Introduction}
\label{sec:intro}




With intelligence marching from the cloud to the edge, autonomous robots are rising in real-world deployment.
They require data to be processed from sensing to actuation in a closed-loop manner. The onboard circuits typically take power from the battery, limiting supported operational time. Recently, with neural networks being deployed at the edge, the demanding real-time performance and energy constraints are becoming increasingly difficult to achieve. Aerial platforms have an additional constraint of payload which limits the weight of the onboard sensors and compute. For tiny terrestrial robots, the form factor becomes a crucial requirement. Therefore, there is a strong need to develop energy-efficient compute platforms and processing frameworks to enable robotic edge intelligence.


Several hardware platforms have been used for robotic applications. CPUs and GPUs are designed to handle a wide range of robotic tasks and algorithms development. However, they usually consume 10-100~W of power, which are orders of magnitude higher than available resources on edge robotic systems. Edge processors like Jetson Nano and Google TPU have been used for programmability at a smaller form factor for faster system-level prototyping~\cite{chinchali2021network}. FPGAs are attracting attention due to its reconfigurability and hardware-efficiency, and have been presented for robotic perception~\cite{gao2021ielas,wan2021energy}, localization~\cite{liu2021archytas,liu2022energyefficient}, and planning~\cite{murray2016microarchitecture,neuman2021robomorphic}. The partial reconfiguration technique takes this flexibility one step further, where part of FPGA resources can be reconfigured at runtime without compromising the other parts of applications~\cite{wan2021survey,liu2021robotic}.

Edge-robotic specialized ASICs boost the energy-efficiency and customization even further. The constrained power budget on edge robots requires these chips to have a few mW of power consumption. 
For robotic perception, Jeon et al.~\cite{jeon_uav} present a feature extraction accelerator at 2.7~mW power for micro aerial vehicles.
For robotic autonomous navigation, Navion~\cite{suleiman2019navion} accelerates the visual-inertial odometry of nano drones, Li et al.~\cite{li2020high} speed up mutual information computation, and Li et al.~\cite{li2019879gops} combine neural network with physical localization model with proposed specialized accelerators.
For end-to-end learning-based robotic control, Kim et al.~\cite{rl_acc} develop a reinforcement learning accelerator for micro drones with 1.1~mW power consumption.
For multi-robots scenarios, Honkote et al.~\cite{search_rescue} design a low-power SoC for distributed and collaborative swarm robot systems.
Different techniques ranging from voltage-mode circuits~\cite{voltage_mode_circuits}, quantized neural networks~\cite{binary_nn,lam2019quantized,tambe2020algorithm} and sparse coding~\cite{sparse_coding} have been utilized in restricting the power consumption.
Several neural network accelerators have been proposed for general vision applications~\cite{cnn_acc1} which can be readily applied to robotic tasks.



Apart from energy consumption, memory bandwidth availability also becomes critical for high-speed edge robotic applications. A newer type of visual sensor called event cameras is being explored for such scenarios. Event cameras provide a stream of asynchronous events, and only part of frames are processed at every time instance, allowing high bandwidth, high speed, and high dynamic range~\cite{dvs_survey}. System-level applications of this include a 3~ms latency robotic goalie \cite{robotic_golie} and looming obstacle avoiding drone~\cite{dvs_drone}. The event-based processing modality of event cameras also lends itself naturally to processing bio-inspired spiking neural networks that benefit from data sparsity and low power.  Tasks like lane detection and prey capturing~\cite{looming_object} have been demonstrated using spiking neural networks and event camera pairs.


In this paper, we will show a series of our proposed ultra-low-power accelerators and system demonstrations for edge robotics, with an emphasis on circuit and system technologies. These demonstrations augment the large landscape of edge robotic systems in reinforcement and swarm learning and neuro-inspired computing technologies. Section~\ref{sec:RL} presents a time-domain mixed-signal accelerator that supports embedded reinforcement learning control. Section~\ref{sec:slam} introduces a hybrid-digital mixed-signal hardware design to enable edge swarm intelligence. Section~\ref{sec:slam} presents an oscillator-based hardware platform for neuro-inspired mapping, localization, and control. Section~\ref{sec:benchmarking} further provides benchmarking and simulation frameworks for cross-layer robotic workload evaluation.
The paper concludes with challenges and opportunities for next-generation hardware-efficient robotic computing (Section~\ref{sec:challenges}).
\section{Reinforcement Learning on the Edge}
\label{sec:RL}
This section presents our reinforcement learning neuromorphic accelerator for robotic autonomous navigation and exploration~\cite{amravati201855nm,amaravati201855}. We propose an energy-efficient time-domain mixed-signal computational framework to enable ultra-low-power operations.

\subsection{Reinforcement Learning Algorithm}
\label{subsec:algo_RL}

Achieving true autonomy requires robots to perform continuous learning through frequently interacting with environments. Reinforcement learning (RL) is such a paradigm where robots take actions in environments to maximize the notion of cumulative reward.
Among all RL algorithms, Q-learning is one of the well-studied techniques, which is implemented in this design. Q-learning seeks to find the optimal action policy ($A_t$) given the current state ($S_t$) to maximizes the reward ($R_t$). It works on the principle of the action-value function $[Q(S_t, A_t)]$, where $Q$ is iteratively updated from Bellman equation. For a detailed overview of RL algorithms and applications, interested readers are referred to~\cite{sutton2018reinforcement}.

\subsection{Circuit and System of RL Neuromorphic Accelerator}
\label{subsec:arch_RL}
\subsubsection{Overview}
\begin{figure}[h]
\vspace{-0.1in}
        \centering\includegraphics[width=\columnwidth]{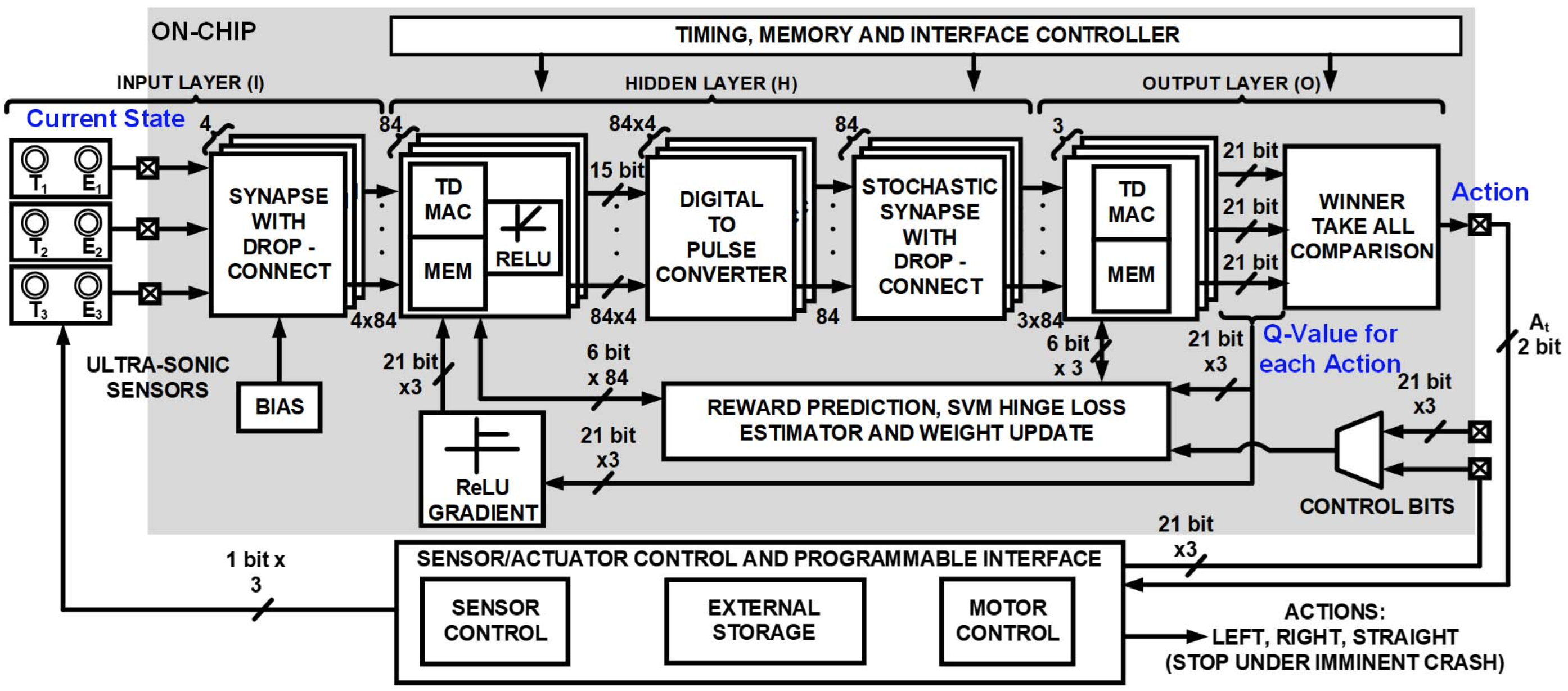}
        \caption{\small The system architecture of our RL neuromorphic accelerator, including different circuit blocks and the interface to the external micro-controllers and motor drivers.}
        \label{fig:RL_system}
        \vspace{-0.1in}
\end{figure}

\ 
\newline
\indent 
Fig.~\ref{fig:RL_system} shows the system diagram of our RL neuromorphic accelerator. It consists of ultrasonic sensors, an RL test chip, a Raspberry Pi-based micro-controller, and motor drivers.

The ultrasonic sensors feed depth information to the input layer of the neural network through an array of stochastic synapses. A three-layer neural network is implemented to process sensor data and generate actions that robots will follow. The micro-controller stores  $(S_t, A_t, R_t, S_{t+1})$ in a scratchpad memory during training and sends action commands from the chip to the motor controllers.

\subsubsection{Time-Domain Mixed-Signal Circuits}
\begin{figure}[h]
        \centering\includegraphics[width=\columnwidth]{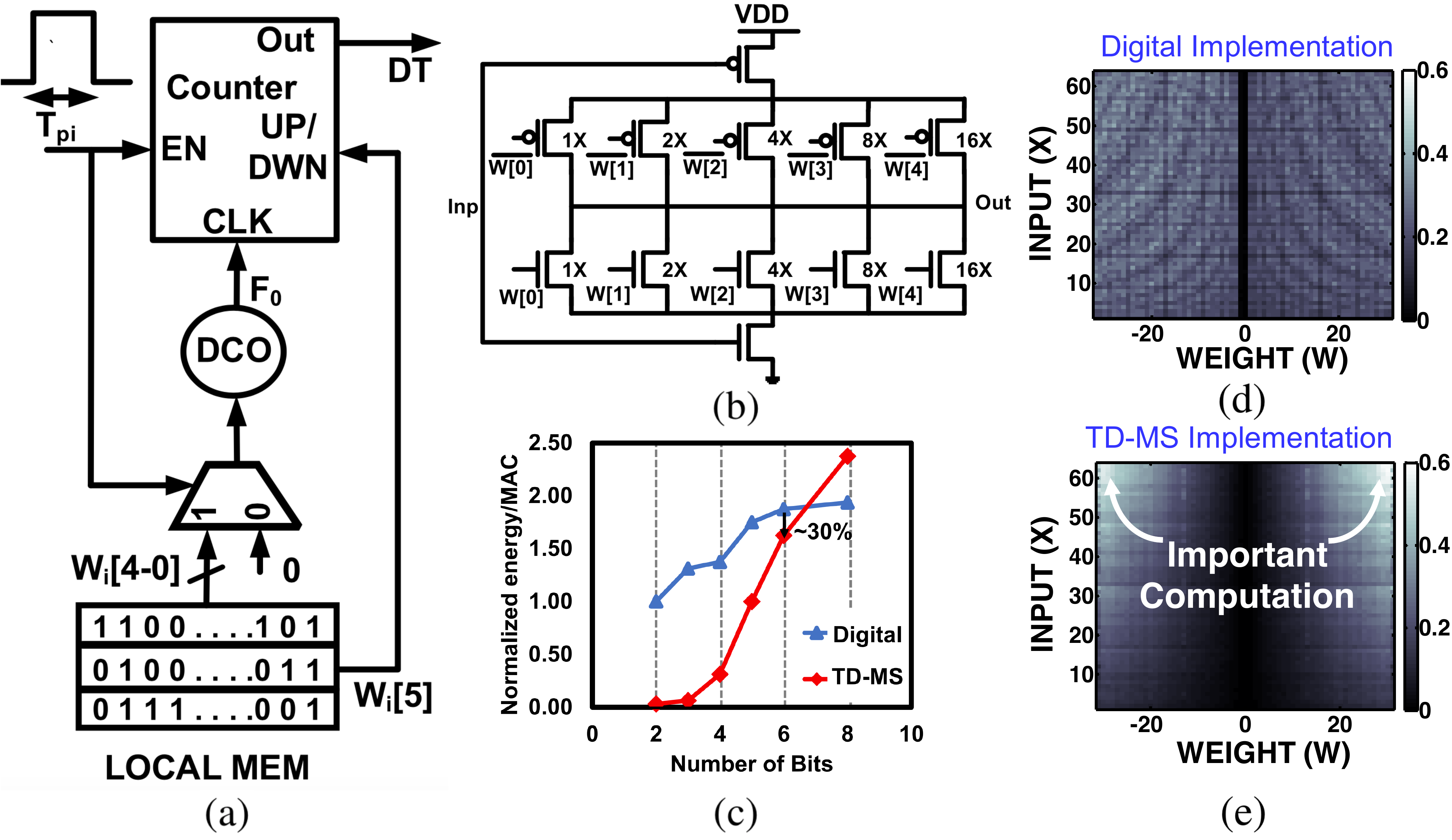}
        \caption{\small (a) Proposed Time-Domain Mixed-Signal (TD-MS) MAC unit. (b) Digitally controlled oscillator (DCO) in TD-MS unit. (c) Computational energy per MAC for digital and TD-MS design with different bit precision at 0.6~V. (d)(e) 2-D energy per MAC surface for digital and TD-MS design with 6-bit inputs at 0.6~V.}
        \label{fig:TDMS}
\end{figure}

\ 
\newline
\indent 
To enable low-power edge intelligence on the robotic platform with low bit precision,
we propose a time-domain mixed-signal (TD-MS) computational framework for energy-efficient and accurate operations. Fig.~\ref{fig:TDMS}\textcolor{blue}{a} illustrates the TD-MS multiply-and-accumulate (MAC) unit. A pulse input from sensor or hidden layers is used to enable the up-down counter that is triggered by a digitally controlled oscillator (DCO) (Fig.~\ref{fig:TDMS}\textcolor{blue}{b}). The output of the counter is the product of weight $W_i$ (digital) and pulse width $T_{pi}$ of the gating signal (analog).

The proposed TD-MS design demonstrates unique advantages: 1) improved energy-efficiency at lower bit width compared to digital MAC operation (Fig.~\ref{fig:TDMS}\textcolor{blue}{c}); (2) the energy to compute is proportional to the significance of the computation - a feature in human brain but missing in digital logic (Fig.~\ref{fig:TDMS}\textcolor{blue}{d-}\ref{fig:TDMS}\textcolor{blue}{e}); (3) 45\% lower system area, 47\% lower interconnect power and 16\% lower leakage power compared to digital design.

\subsubsection{Enabling Regularization via Stochasticity}
\begin{figure}[h]
        \centering\includegraphics[width=\columnwidth]{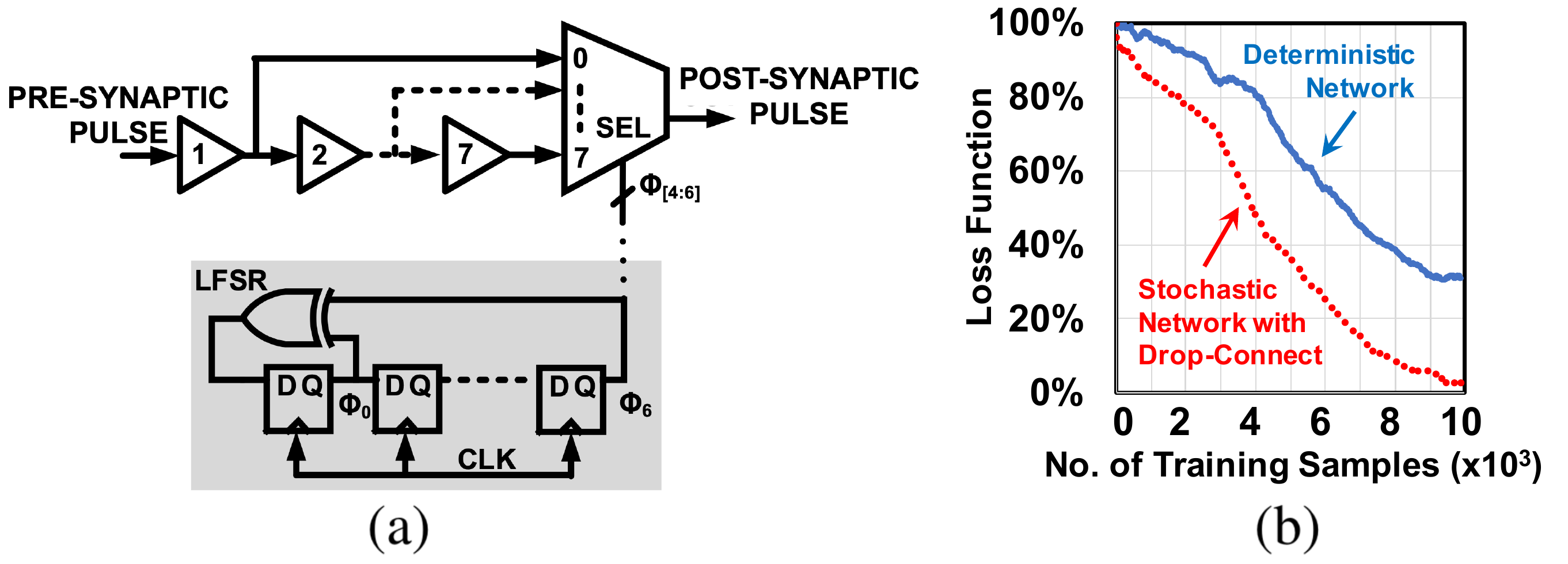}
        \caption{\small (a) Stochasticity is implemented by introducing varying delays between bit transitions using the linear-feedback-shift register (LFSR). (b) Stochasticity accelerates policy convergence.}
        \label{fig:synapse}
\end{figure}

\ 
\newline
\indent 
To help generalize the learned neural network model to unknown environments, we introduce stochasticity to synapses with drop-connect. As shown in Fig.~\ref{fig:synapse}\textcolor{blue}{a}, the stochasticity is implemented with a buffer chain whose delays are randomly altered by a local high-speed linear-feedback-shift register (LFSR). Compared to the deterministic model, the stochastic network can achieve 1.7$\times$ speedup in convergence (Fig.~\ref{fig:synapse}\textcolor{blue}{b}).

\subsection{Evaluation}
\label{subsec:eval_RL}
\begin{figure}[h]
        \centering\includegraphics[width=\columnwidth]{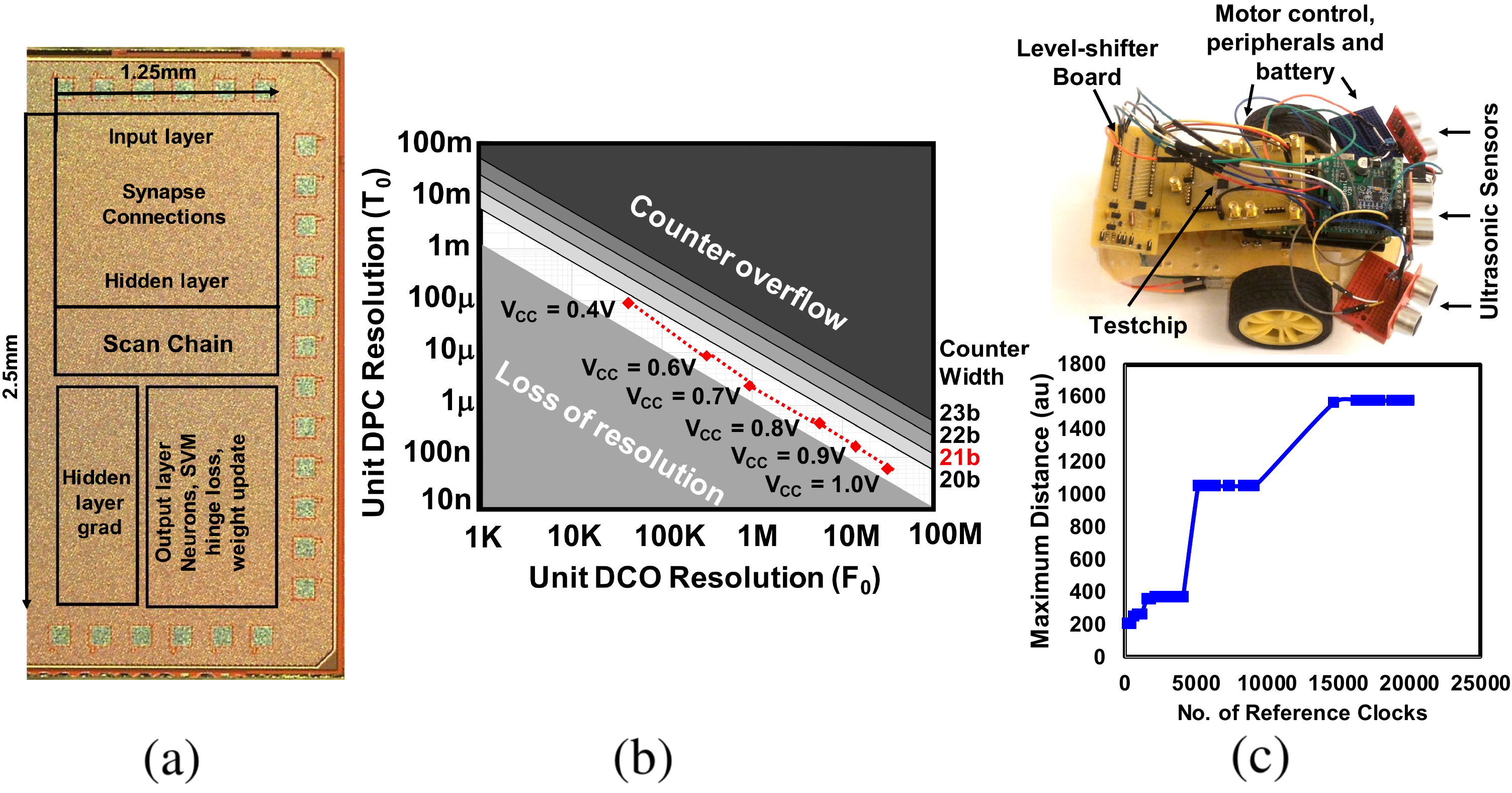}
        \caption{\small (a) RL neuromorphic chip die-photo. (b) Design space showing scalability till 0.4~V. (c) Mobile robot system and its covered distance as a function of the number of clock cycles or iterations. }
        \label{fig:RL_result}
        \vspace{-0.15in}
\end{figure}

The RL neuromorphic test chip is implemented and taped-out in 55~nm CMOS process (Fig.~\ref{fig:RL_result}\textcolor{blue}{a}). By exploring design space of proposed TD-MS circuits, we show that the chip can ensure correct functionality and voltage scalability from 1.0~V down to 0.4~V (Fig.~\ref{fig:RL_result}\textcolor{blue}{b}). The chip is mounted on a tiny mobile robot for autonomous exploration and learning. Fig.~\ref{fig:RL_result}\textcolor{blue}{c} illustrates the increasing moved distance when robot continuously explores the space in the presence of obstacles. 

\section{Swarm Intelligence on the Edge}
\label{sec:swarm}
Going beyond single tasks, when facing a variety of tasks, swarms of robots usually collaborate with each other to solve problems. This section presents our energy-efficient hardware accelerator that enables swarm robotic applications~\cite{cao201914,cao201965}.

\subsection{Swarm Intelligence Algorithms}
\label{subsec:swarm_algo}
Swarm algorithms can be broadly classified into two categories, ones based on physical and mathematical models (model-based) and ones based on learning (model-free).
Among model-based algorithms, artificial potential field (APF) is a commonly-used approach for collaborative path planning, where the motion force is obtained by aggregating the attractive and repulsive potential field. Model-free algorithms usually allow each robot to learn continuously to establish a model with real-world knowledge without human intervention. RL-based cooperative algorithms have shown great promise. Interested readers are directed to~\cite{la2014multirobot} for more details.

\begin{figure}[h]
        \centering\includegraphics[width=\columnwidth]{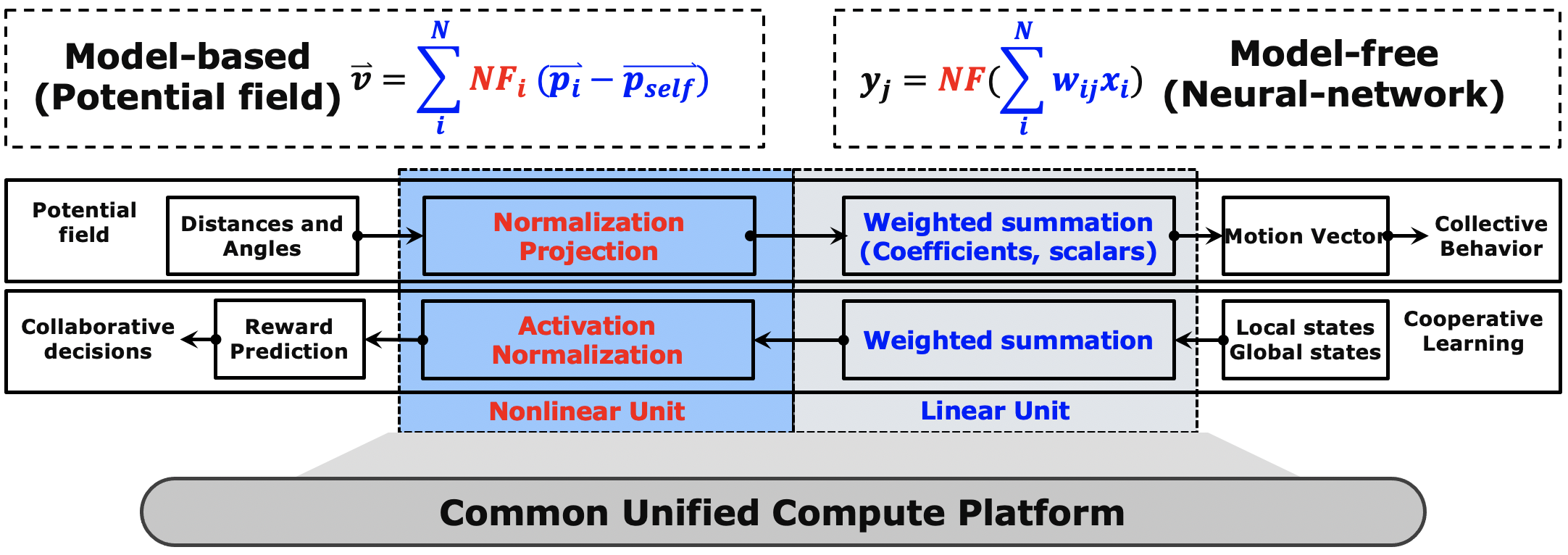}
        \caption{\small Common unified compute paradigm that supports both model-based and model-free swarm algorithms.}
        \label{fig:unified_paradigm}
        \vspace{-0.1in}
\end{figure}

Interestingly, we observe that both model-based and model-free autonomy paradigms have similar mathematical structures (Fig.~\ref{fig:unified_paradigm}).
Therefore, we identify the commonalities and develop a unified architecture to support real-time swarm intelligence.

\subsection{Circuit and System of Swarm Intelligence Accelerator}
\subsubsection{Overview}
\begin{figure}[h]
        \centering\includegraphics[width=.7\columnwidth]{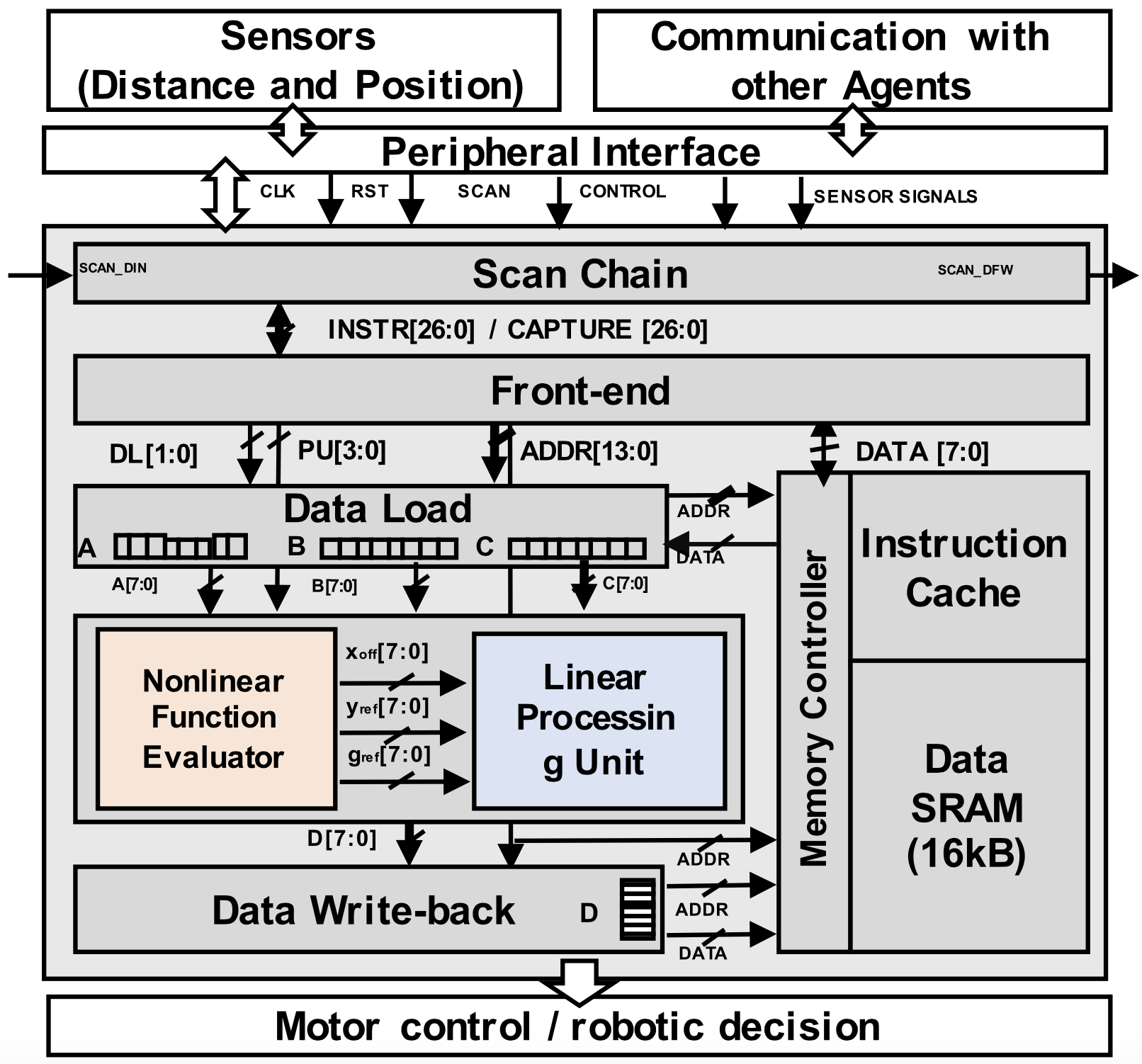}
        \caption{\small The system architecture of our swarm intelligence accelerator, supporting both swarm-based and swarm-free autonomy paradigms with a nonlinear function evaluator and linear processing unit. }
        \label{fig:swarm_arch}
\end{figure}

\ 
\newline
\indent 
The system architecture of our swarm intelligence accelerator is shown in Fig.~\ref{fig:swarm_arch}. Noticing that both model-based and model-free are combinations of nonlinear and linear operations, we design a nonlinear function evaluator (NFE) and linear processing unit (LPU). 
NFE supports nonlinear operations by using a piecewise linear approximation of nonlinear functions.
LPU supports all addition and multiplication linear operations. Most operations are implemented in the digital domain except for MAC that leverages mixed-signal computing.
The datapath of NFE and LPU is bi-directional, so data can move between each other seamlessly and preserve locality.
We also observe that several required functions show symmetry and periodicity, which provides a further chance to reduce the number of computations and comparisons.

\subsubsection{Hybrid-Digital Mixed-Signal Circuits Architecture}
\ 
\newline
\indent 
Swarm algorithms need to support various swarm sizes in dynamic environments. The required bit precision will increase from 3-bit to 8-bit, with swarm size increasing from 2 agents to 20 agents. TD-MS MACs show energy advantages over digital counterparts for low bit width, but exhibit higher energy with increasing operand size (Fig.~\ref{fig:TDMS}\textcolor{blue}{c}). To address this issue, we propose a hybrid-digital mixed-signal (HD-MS) MAC kernel, where computation is purely TD-MS for bit-width$\leq$5, and hybrid of TD-MS and digital for 6$\leq$bit-width$\leq$8. 

\begin{figure}[h]
        \centering\includegraphics[width=\columnwidth]{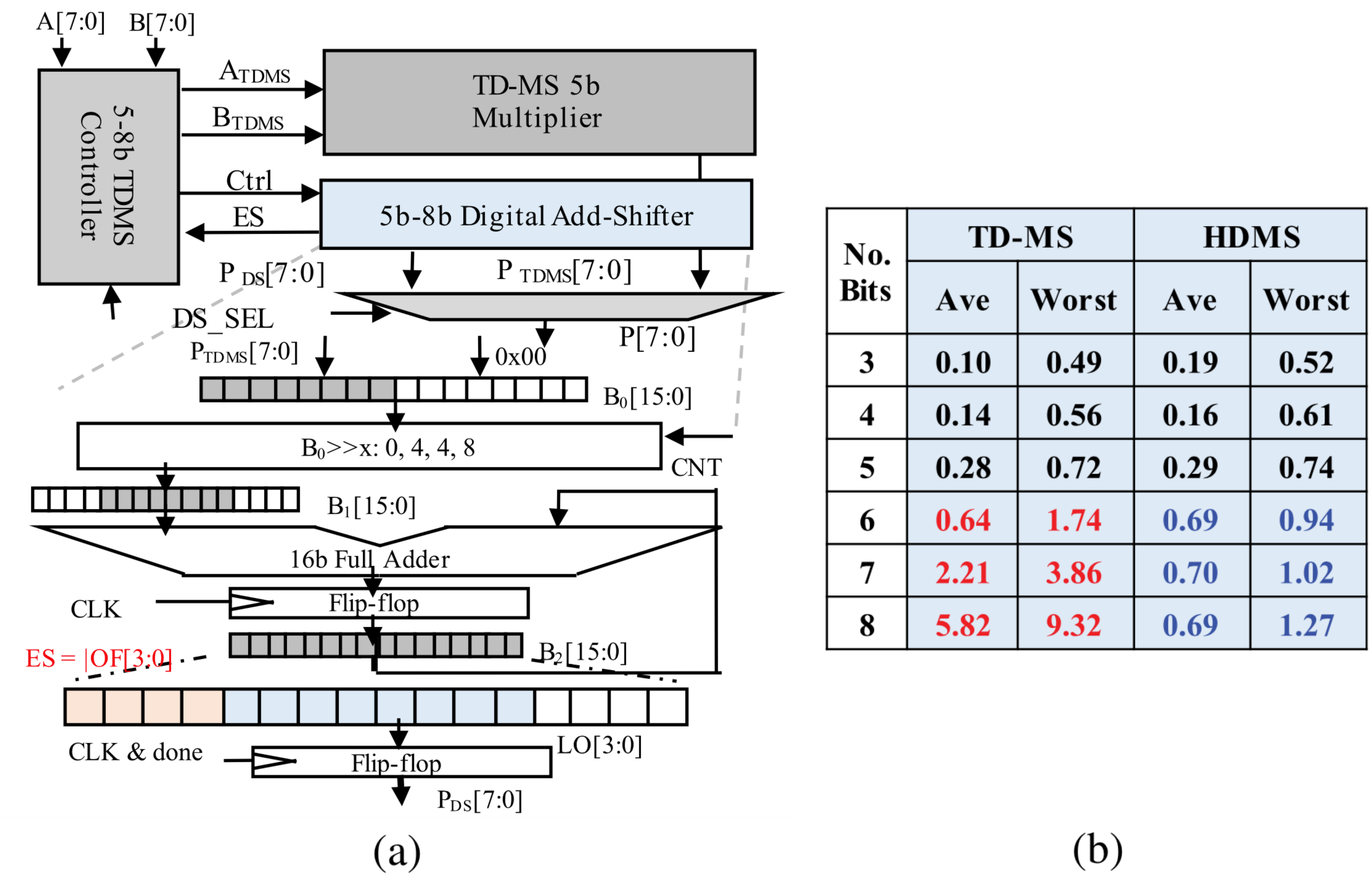}
        \caption{\small (a) Circuit schematic of the HD-MS design, including the 5-bit TD-MS kernel and the digital peripherals to enable efficient scaling to 8-bit. (b) Energy/MAC (normalized to a digital implementation) for TD-MS and HD-MS implementations. We observe that HD-MS outperforms TD-MS and digital for large swarm sizes.}
        \label{fig:swarm_hdms}
        \vspace{-0.1in}
\end{figure}

Fig.~\ref{fig:swarm_hdms}\textcolor{blue}{a} shows the circuit schematic of the HD-MS design. The HD-MS MAC kernel consists of a conventional TD-MS  multiplier, a 5-8-bit TD-MS controller, and a 5-8-bit digital adder-shifter. The TD-MS multiplier computes $\leq$5-bit operation. The TD-MS controller and digital adder-shifter reconfigure the multiplier to a higher bit width with seamless shift-and-add operations. Fig.~\ref{fig:swarm_hdms}\textcolor{blue}{b} demonstrates that HD-MS have energy-benefits at higher precision compared with TD-MS. Compared with digital implementation, HD-MS exhibits 81\% (for 3-bit) to 31\% (for 8-bit) energy per MAC reduction.

\subsection{Evaluation}
\label{subsec:swarm_testchip}
\begin{figure}[h]
        \centering\includegraphics[width=\columnwidth]{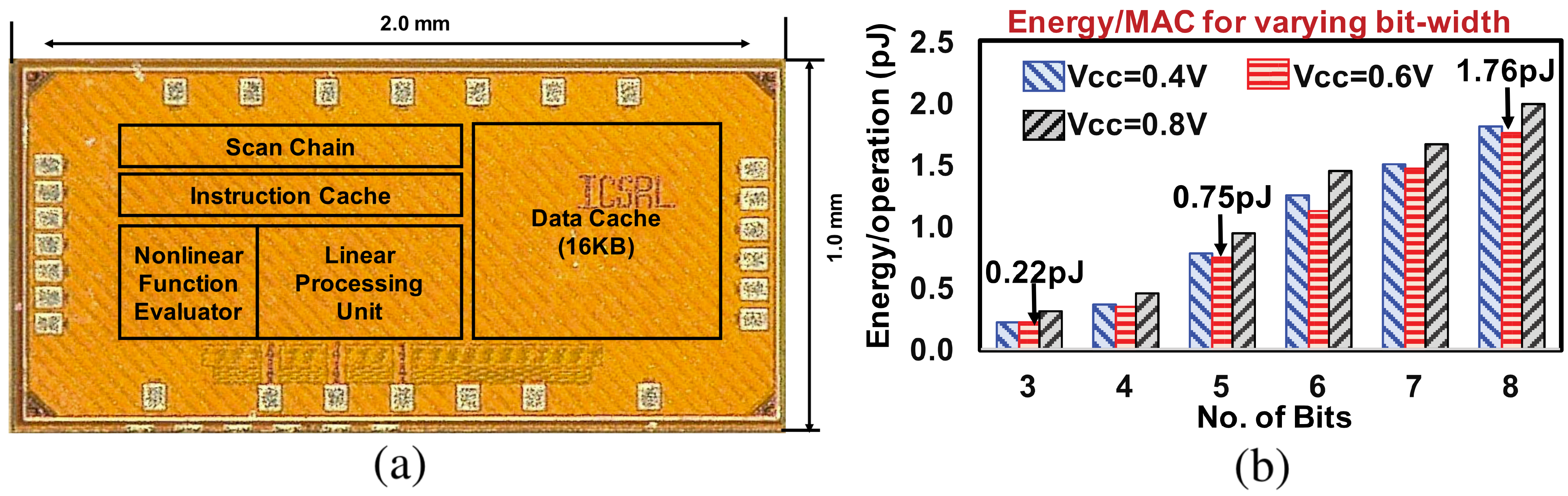}
        \caption{\small (a) Swarm intelligence chip die-photo. (b) Measured energy per MAC across different bit widths at VCC = 0.4, 0.6, and 0.8 V.}
        \label{fig:swarm_testchip}
        \vspace{-0.15in}
\end{figure}

Our swarm intelligence accelerator is fabricated in 65~nm CMOS process (Fig.~\ref{fig:swarm_testchip}\textcolor{blue}{a}). Fig.~\ref{fig:swarm_testchip}\textcolor{blue}{b} demonstrates its scalability with bit precision, indicating a peak energy efficiency of 0.22~pJ/MAC (for 3-bit) and 1.76~pJ/MAC (for 8-bit). The average energy efficiency varies from 9.1 TOPS/W (for 3-bit) to 1.1~TOPS/W (for 8-bit). This bit precision scalability allows efficient computation for various swarm sizes.

\begin{figure}[h]
\vspace{-0.1in}
        \centering\includegraphics[width=\columnwidth]{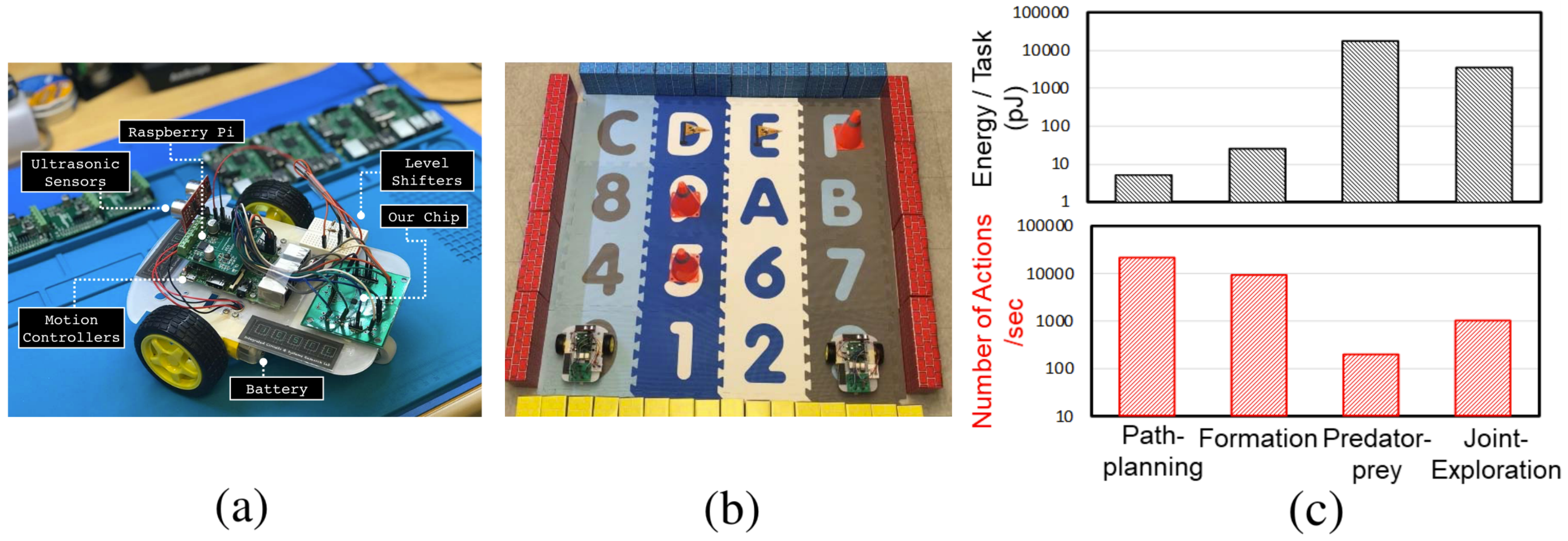}
        \caption{\small (a) Our swarm intelligence accelerator is mounted on a robotic car with peripheral circuits. (b) Experimental setup. (c) Energy and performance for different template swarm algorithms.}
        \label{fig:swarm_demo}
        \vspace{-0.15in}
\end{figure}

We mounted the chip on a robotic car (Fig.~\ref{fig:swarm_demo}\textcolor{blue}{a} and Fig.~\ref{fig:swarm_demo}\textcolor{blue}{b}). The platform interfaces with a Raspberry Pi, sensors, motor controllers, and radios. We implement four exampled swarm intelligence algorithms, namely path planning, pattern formation, predator-prey, and joint-exploration, where the first two are model-based and the last two are model-free. Fig.~\ref{fig:swarm_demo}\textcolor{blue}{c} shows large variations in energy consumption and the number of actions for each task, illustrating that future robotic platforms need to support a wide variety of algorithms, as the complexities of environments and task can dramatically change.

\section{Neuro-inspired Computing on the Edge}
\label{sec:slam}
Recently, spiking neural networks (SNNs), with bio-inspired sparse spike-based temporally coded computing, offer an energy-efficient alternative for edge robotic tasks. This section presents our proposed incorporation of SNNs for SLAM and prey tasks for ultra-low-power robotic applications~\cite{yoon202031,yoon2020neuroslam}.

\subsection{Simultaneous Localization and Mapping Algorithms}
\label{subsec:slam_algo}
\begin{figure}[h]
\vspace{-0.1in}
        \centering\includegraphics[width=.75\columnwidth]{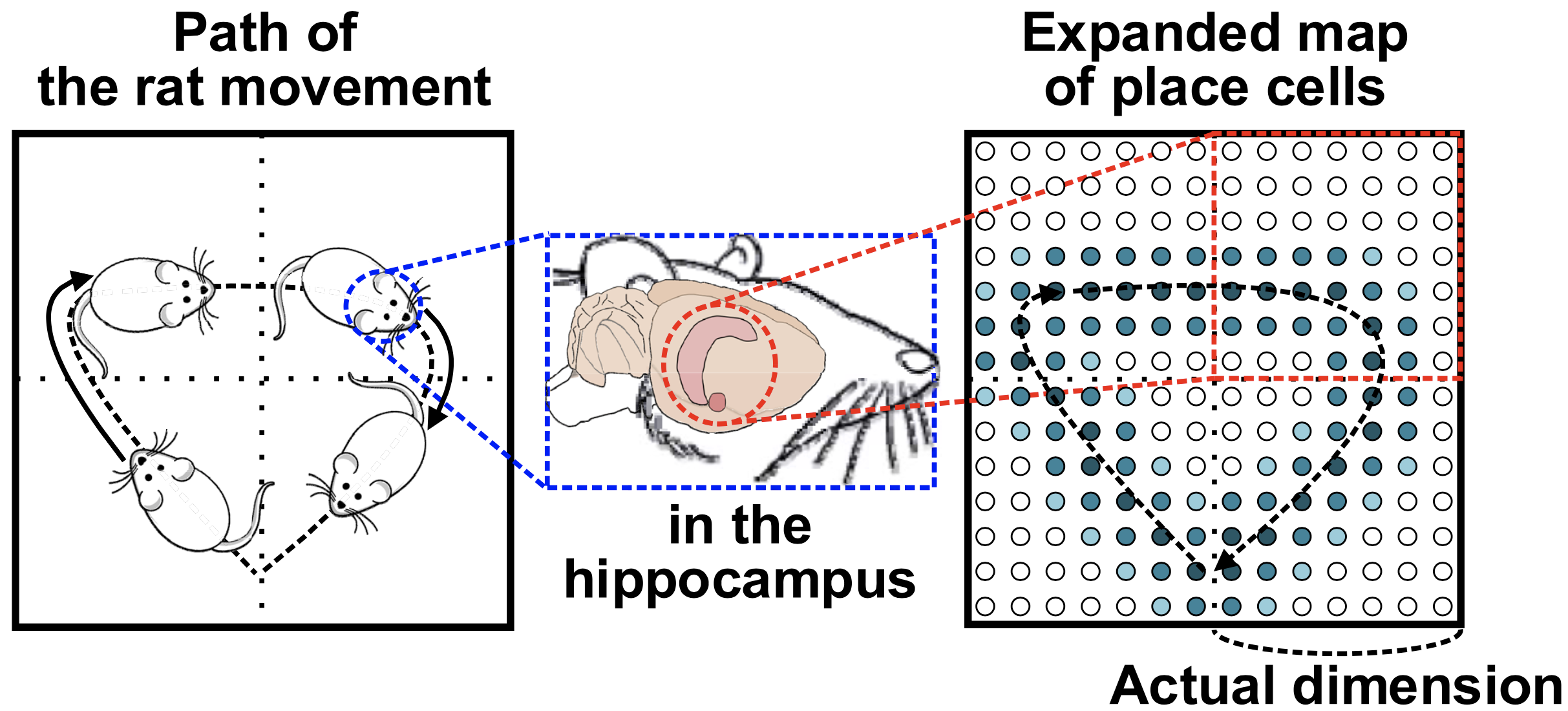}
        \caption{\small Mapping between the position of rodent and the excited place cells in rodent's brain}
        \label{fig:ratslam_mapping}
        \vspace{-0.1in}
\end{figure}

Simultaneous localization and mapping (SLAM) forms an essential component of many autonomous navigation applications. SLAM  algorithms can be broadly classified into two categories: visual-based and neuro-based.
Visual-based SLAM requires the robot to identify its position from the beginning of motion and generate a map of the movement using only the images captured during motion. Previous approaches like probabilistic SLAM and keyframe-based SLAM remain inadequate due to the constrained power budget of edge systems. 
Neuro-based SLAM performs the computation in a more energy-efficient way and is applicable to ultra-low-power edge robotics. Neuroscientific exploration in rodent brains showed their phenomenal capacity to efficiently localize themselves (Fig.~\ref{fig:ratslam_mapping}). Place cells and head direction cells are identified as neuronal circuits tuned to respond at a particular position and direction of motion, respectively. Our NeuroSLAM accelerator takes inspiration from RatSLAM~\cite{ratslam} algorithm that mimics the neuromorphic connectivity and incorporates bio-inspired hardware to achieve ultra-low-power SLAM tasks.

\subsection{Circuit and System of NeuroSLAM Accelerator}
\subsubsection{Architecture and Circuit Block}
\begin{figure}[h]
        \centering\includegraphics[width=.92\columnwidth]{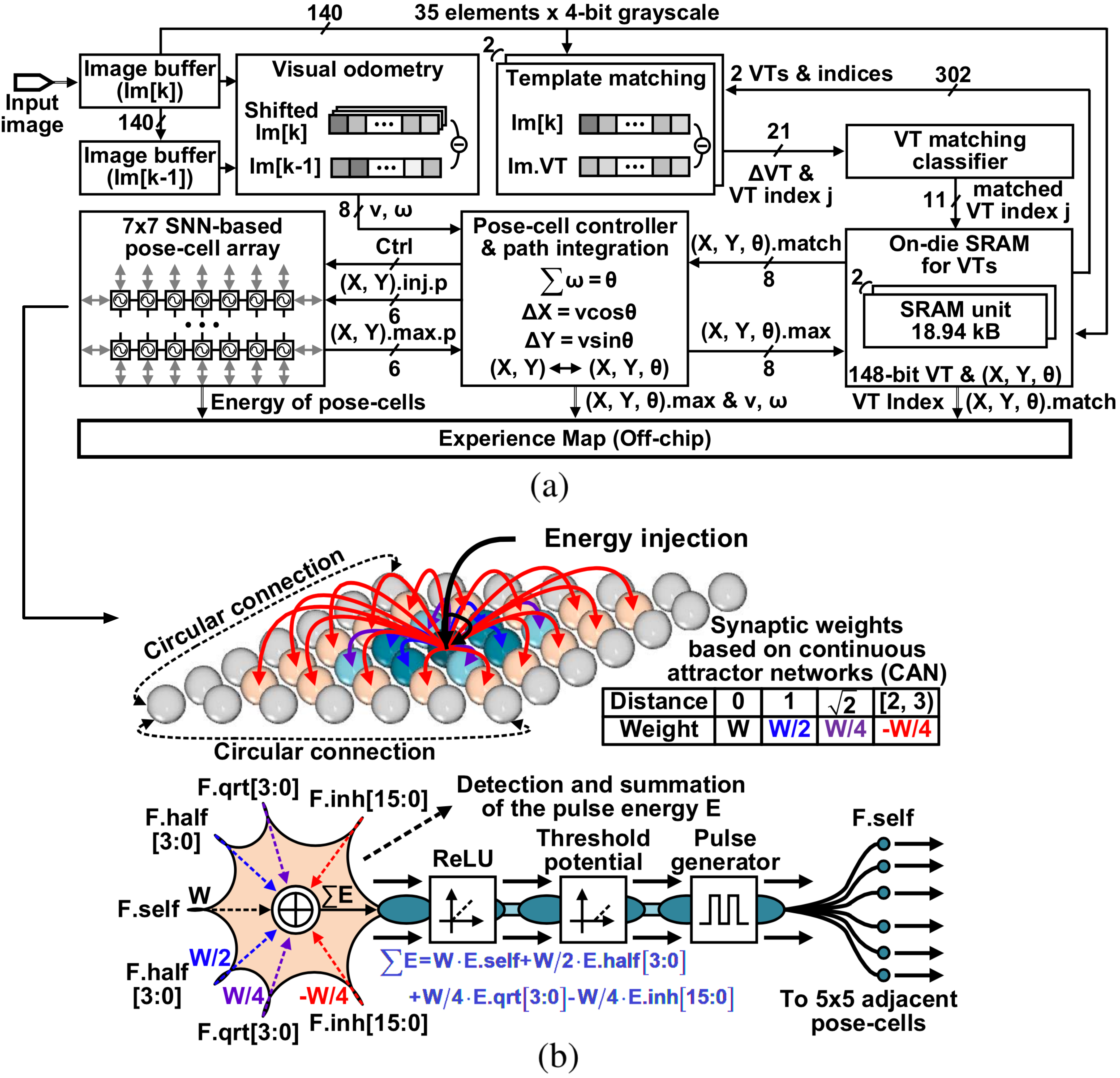}
        \caption{\small (a) Overview of our NeuroSLAM accelerator. (b) Spiking continuous attractor network mimicking rodent's pose cell behavior.}
        \label{fig:neuroslam_intro}
        \vspace{-0.15in}
\end{figure}

\ 
\newline
\indent 
The flow of data in the NeuroSLAM accelerator is shown in Fig.~\ref{fig:neuroslam_intro}\textcolor{blue}{a}, where the captured image is first compared with the previous image for visual odometry to estimate the displacement from previous image capture. This is followed by template matching with previous images to detect any possibility of loop closure. The translation is added with the previous position to find the current direction of motion with digital head direction cells, and the path integration is injected in the pose cell array. Pose cell array is made of oscillator-based continuous attractor network in spiking neural network fashion, as shown in Fig.~\ref{fig:neuroslam_intro}\textcolor{blue}{b}. The output is extracted to calculate the experience map.

\label{subsec:slam_algo}

\subsubsection{Test Chip Measurement}
\begin{figure}[h]
        \centering\includegraphics[width=.9\columnwidth]{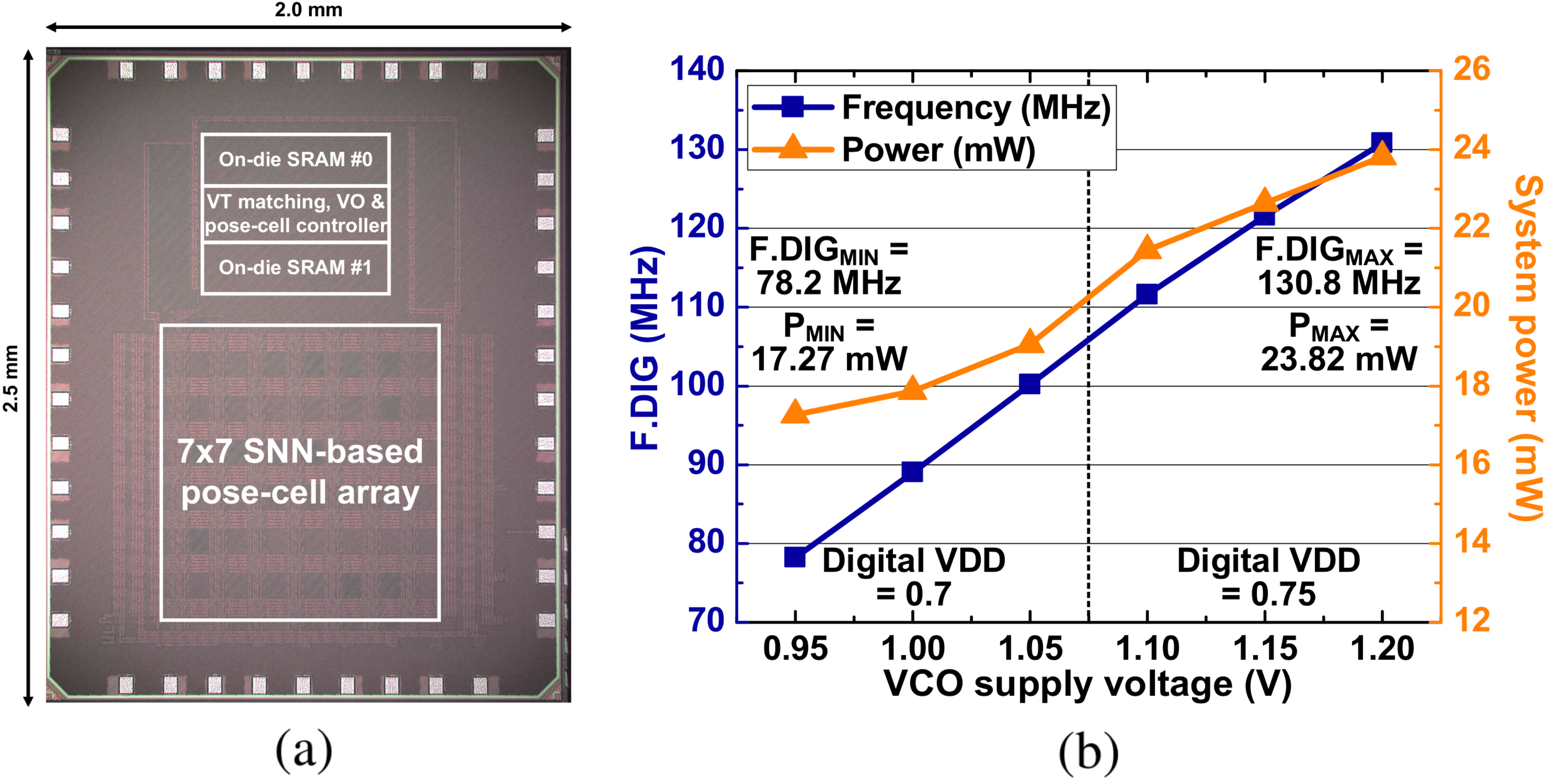}
        \caption{\small(a) NeuroSLAM chip die-photo. (b) Measured operational frequency and power consumption of NeuroSLAM accelerator.}
        \label{fig:neuroslam_results}
        \vspace{-0.2in}
\end{figure}

\ 
\newline
\indent 
The NeuroSLAM test chip is fabricated in 65~nm technology (Fig.~\ref{fig:neuroslam_results}\textcolor{blue}{a}). The template matching, odometry, and path integration are carried out in low-bit resolution with an efficient attractor network. The chip achieves SLAM with only 23.82~mW. The dependence of power consumption on input voltage is shown in Fig.~\ref{fig:neuroslam_results}\textcolor{blue}{b}. The attractor network shows a high compute efficiency of 8.79 TOPS/W. Our NeuroSLAM accelerator demonstrates the incorporation of neuro-inspired analog hardware in the digital pipeline to enable edge intelligence in severe energy-constrained systems.

\subsection{Neuro-Inspired End-to-End Spike-Only Processing}
\begin{figure}[t!]
        \centering\includegraphics[width=.75\columnwidth]{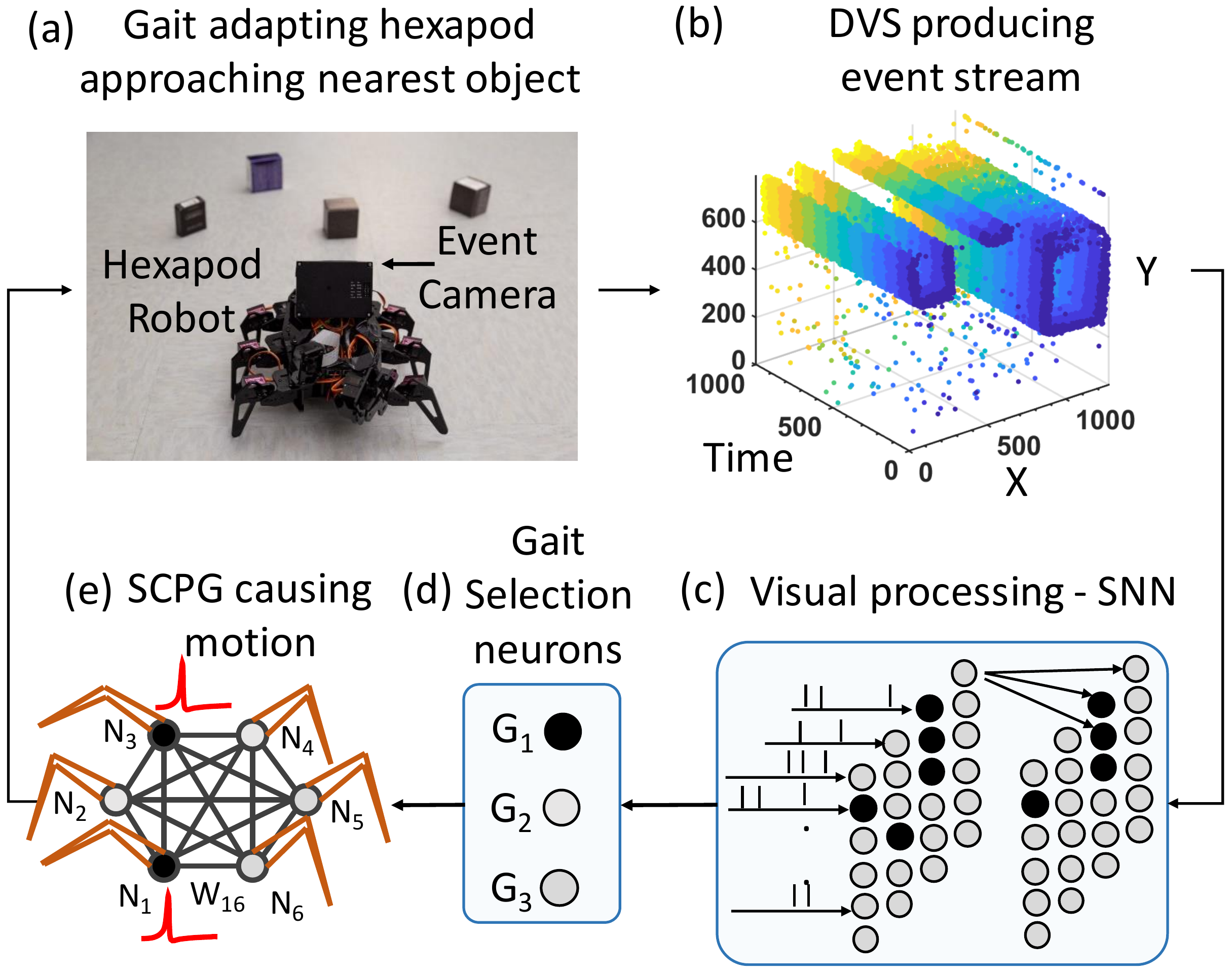}
        \vspace{-0.07in}
        \caption{\small Bio-inspired closed-loop end-to-end spike-only robot. (a) Event camera mounted on hexapod provides (b) input stream to (c) SNN for processing and (d) selecting gait to approach the nearest object. (e) Gait is executed by a spiking CPG causing leg movements.}
        \label{fig:snn_cl}
        \vspace{-0.25in}
\end{figure}

Another neuro-inspired system exploration is the first autonomous sensing-to-actuation end-to-end spike-only processing pipeline for hexapod robots~\cite{lele2021end}. The goal is to demonstrate the functionality of spike-only processing and evaluate the potential of event-driven processing modalities. As shown in Fig.~\ref{fig:snn_cl}, event camera/dynamic vision sensor (DVS) is used as the sensory input to generate asynchronous event stream. The information is processed through SNN to activate one of the three gait selection neurons. The central pattern generator (CPG) is trained such that every gait selection neuron activates gait in a different direction to allow controlled movement. A task of identifying and approaching the nearest target is demonstrated using this platform. The ultra-low-power (2.55~mJ/step) consumption of this system with dedicated spiking hardware (e.g., Loihi) highlights the potential of neuro-inspired event-driven systems for edge applications. 

\section{Benchmarking and Software Infrastructure}
\label{sec:benchmarking}
Edge robotics is a cross-layer research field, spanning from environment modeling, autonomy algorithms, runtime systems to onboard compute architecture and circuits. The interaction between the layers impacts the efficacy and performance of the system~\cite{krishnan2020sky}, requiring software infrastructure for interdisciplinary research. Specially, we need a platform that can systematically benchmark each of these individual layers and also capture end-to-end cross-layer execution characteristics. Recently, some platforms for aerial robots have been proposed. MAVBench~\cite{boroujerdian2018mavbench} is a platform for physical-model-based aerial robots evaluation. MAVBench involves a closed-loop simulator and a benchmark suite for several computational kernels involving perception, planning, and control. Airlearning~\cite{krishnan2019air} and PEDRA~\cite{anwar2018navren} are simulation suits and benchmarks for learning-based aerial edge robots. They provide a rich set of virtual worlds, including indoor and outdoor environments, to enable autonomy generalization. In addition, PEDRA supports swarm intelligence with different collaboration paradigms.

These benchmarking and software infrastructures inspire a proliferation of studies on edge robotic systems, such as reliability, system design methodology, and memory hierarchy optimization. 
Based on MAVBench, RoboRun~\cite{boroujerdian2021roborun} presents a robot runtime that leverages spatial-aware computing to dynamically improve performance and energy for heterogeneous operating environments. MAVFI~\cite{hsiao2021mavfi} proposes a fault injection framework for end-to-end reliability characterization of robotic workload, which is portable to robot operating system (ROS)-based applications.
Based on AirLearning, Skyline~\cite{wan2021roofline}, Autopilot~\cite{krishnan2021machine}, and AutoSoC~\cite{krishnan2021autosoc} propose a visual performance model and automated design space exploration framework to design optimal onboard compute for aerial robots. 
Based on PEDRA, Anwar et al.~\cite{anwar2020autonomous} present a transfer learning-based approach to reduce the onboard computation required to train a neural network for autonomous navigation. Wan et al.~\cite{wan2021analyzing} and FRL-FI~\cite{wan2022frlfi} propose application-aware lightweight fault detection and mitigation techniques to enable reliable autonomy under hardware faults, for both learning-based and swarm-based intelligence. 
Zeng et al.~\cite{zeng2021decentralized} develop a mathematical framework for solving multi-task reinforcement learning problems based on policy gradient method.
Anwar et al.~\cite{anwar2021multi} evaluate the robustness of swarm robotic systems under adversaries. Yoon et al.~\cite{yoon2019hierarchical} present a novel hierarchically memory system with STT-MRAM and SRAM to support real-time learning-based robotic exploration. We believe a holistic benchmarking and simulator infrastructure will uncover more cross-layer research findings of various fields of edge robotics.


\section{Challenges and Opportunities}
\label{sec:challenges}

Robotic computing is a rising area and opens many research challenges and opportunities.
At the device and circuit level, embedded non-volatile memory (e.g., RRAM, Ferroelectric memories) and monolithic 3D integration will provide opportunities for high-performance and energy-efficient robotic computing. 
At the architecture level, the robotic computing platform needs to be adaptive and reconfigurable to various scenarios and support a diverse set of applications, including both DNNs and SNNs. 
At the system level, a holistic benchmarking suite and a generic framework for mapping autonomy algorithms to heterogeneous hardware platforms will benefit the robotic computing development process.
At the algorithm level, robots need to have the ability of lifelong learning and learning with sparse and limited data. The booming of swarm intelligence requires more effective and robust distributed learning across multiple agents.




\section{Conclusion}
\label{sec:conclusion}
This paper investigates circuit and system technologies for energy-efficient edge robotics. We present three robotic accelerators for edge reinforcement learning, swarm intelligence, and neuron-inspired mapping and localization, with an emphasis on novel mixed-signal circuit technologies. We summarize robotic benchmarking and simulation frameworks along with how they foster research efforts in cross-layer robotic systems. Finally, we discuss the challenges and opportunities for the next-generation energy-efficient robotic computing platforms.
\section*{Acknowledgements}
The authors would like to thank Anvesha Amaravati, Saad Bin Nasir, Insik Yoon, Ningyuan Cao, Muya Chang, Jong-Hyeok Yoon, Aqeel Anwar Malik, and Justin Ting for their technical support. This work was supported in part by C-BRIC, one of six centers in JUMP, a Semiconductor Research Corporation (SRC) program sponsored by DARPA. 
\bibliographystyle{ieeetr}
\bibliography{refs}

\end{document}